\documentclass[preprint,preprintnumbers,amsmath,amssymb,superscriptaddress]{revtex4-1}

\usepackage[dvips]{graphicx}
\usepackage{amsmath,bm}
\usepackage{comment}

\usepackage{tikz}
\usetikzlibrary{positioning}

\usepackage{ulem}
\usepackage{color}

\newcommand{\non}{\nonumber}

\begin{document}


\title{Inference Methods for Interaction and Noise Intensities Using Only Spike-time Data on Coupled Oscillators}


\author{Fumito Mori}
 \email{mori@design.kyushu-u.ac.jp}
\affiliation{{Faculty of Design, Kyushu University,  Fukuoka 815-8540, Japan}}
\affiliation{{Education and Research Center for Mathematical and Data Science, Kyushu University,  Fukuoka 815-8540, Japan}}
\author{Hiroshi Kori}
 \email{kori@k.u-tokyo.ac.jp}
\affiliation{{Department of Complexity Science and Engineering, University of Tokyo, Chiba 277-8561, Japan}}


\date{\today}

\begin{abstract}
We propose theoretical methods to infer coupling strength and noise intensity simultaneously through an observation of spike timing in two well-synchronized noisy oscillators.
A phase oscillator model is applied to derive formulae relating each of the parameters to some statistics from spike-time data.
Using these formulae, each parameter is inferred from a specific set of statistics.
We demonstrate the methods with the FitzHugh-Nagumo model as well as the phase  model.
Our methods do not require any external perturbation and thus are ready for application to various experimental systems.
\end{abstract}


\maketitle


Coupled oscillators such as cardiac myocytes \cite{yamauchi02}, heart pacemakers \cite{winfree01,glass01},
 circadian clocks \cite{winfree01,glass01,reppert02},
electro-chemical oscillators \cite{kiss07}, spin torque oscillators \cite{rippard05,kaka05,mancoff05,keller09},
 crystal oscillators \cite{zhou08},
  and nanomechanical oscillators \cite{matheny13} are found in many disciplines ranging from biology to engineering.
Although these systems are subject to various types of noise, including thermal, quantum, and molecular noise, they can exhibit synchronization because of coupling between the oscillators.
Thus, coupling and noise are crucial factors in the determination of multi-oscillator dynamics.

Since a noninvasive estimation 
is desired in many cases,
it is important to develop methods to infer the coupling strength and noise intensity solely
 from temporal information on the oscillation.
Such an attempt was made
in an experiment with cultured cardiac myocytes beating spontaneously \cite{yamauchi02}.
Therein, the transition from a desynchronized state to a synchronized state between two cells was observed within the incubation time.
This suggests that coupling between the cells
should
increase.
However, this naive expectation is not generally fulfilled, because synchronization is facilitated not only by increased coupling strength but also by decreased noise intensity \cite{kuramoto84}.

Figure~1(a) displays spike-time data generated with the FitzHugh-Nagumo model for cardiac and neural electrical activity 
{\cite{FN1961}} {(precisely introduced later)}.
For parameter sets i and ii, the typical values  $\zeta$
of the spike-time lag, which represent the degree of synchronization, are approximately equal.
From this, the coupling strengths in the two cases may seem similar.
However,
the values actually differ by a factor of two.
Thus, an individual statistic derived from oscillation data can be misleading when attempting to infer coupling strength.
The case of attempting to infer noise intensity is similar.
Hence, in order to infer these properties, different types of statistics must be combined appropriately.

In this paper, we propose two methods to infer coupling strength and noise intensity from data solely on the spike timing of two well-synchronized noisy oscillators.
Method I requires spike timing data on only one of the oscillators, but we may infer the coupling strength as well as the noise intensity.
Method II requires spike-time data on both oscillators but provides more precise inferences.
We demonstrate our methods with a phase oscillator model and the FitzHugh-Nagumo model.
An example of our inferences from the FitzHugh-Nagumo model is shown in Fig.~1(b).
There, the coefficient of variation in periods (1.9\% to 4.4\%)
and the number of observed spikes (160,000)
 were comparable to
those in the abovementioned experiment on cardiac cells~\cite{yamauchi02}; i.e., the demonstration in the figure is realistic.
%
While many inference methods work effectively with 
data taken from unsychronized oscillators \cite{tokuda07,miyazaki06,rosenblum01}, 
external perturbation \cite{galan05,Timme07,ota09wsta}, 
or whole time series \cite{tokuda07,miyazaki06,rosenblum01},  
 our methods do not require them.  
 Moreover, we do not need to assume function form. 
  Therefore, our methods are ready for application to synchronized coupled oscillator systems in various fields.

\begin{figure}[h]
\begin{center}
\includegraphics[width=8.5cm]{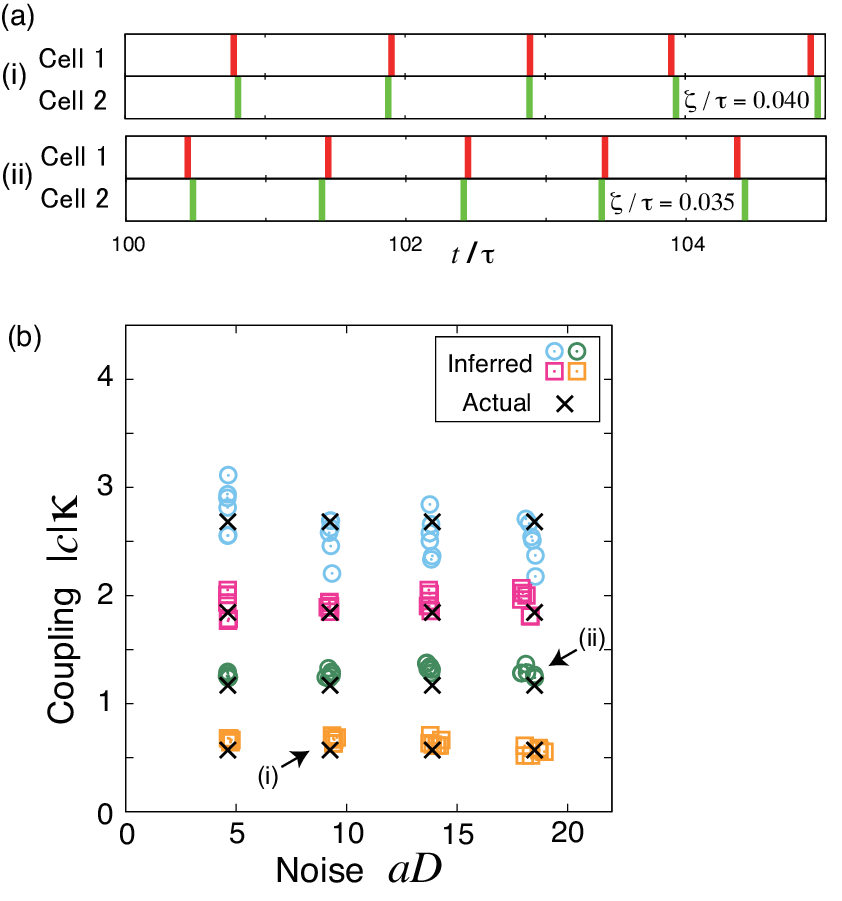}
\end{center}
\caption{(color online)
(a) Examples of spike timing generated for coupled cells with the FitzHugh-Nagumo model in Eq.~(\ref{eq:FN}).
The standard deviation $\zeta$ of spike-time lag between two oscillators are similar in cases i and ii.
(b) Simultaneous inferences of effective noise intensity $aD$
 and effective coupling strength $|c| \kappa$ for the FitzHugh-Nagumo model with method II.
These inferences were achieved using only spike-time data.
Actual values are plotted as crosses.
Inferred values are plotted as squares (for the lowest and third-lowest coupling strength) and
circles
 (otherwise).
The actual values are very well approximated
in all cases, including
(i) $D=0.002$, $\kappa=0.01$ and (ii) $D=0.004$, $\kappa=0.02$.}
\label{fig:fig1}
\end{figure}

We introduce some statistical quantities
based on the spike time data (Fig.~\ref{fig:fig1} (a)).
We assume that an oscillator spikes
when its oscillatory variable passes a specific value.
Let us denote the $k$th spike time of the oscillator by $t^{(k)}$.
In the case of the phase oscillator model, $t^{(k)}$ is defined as the time at which a phase first passes through
$2\pi k + \theta_{\text{cp}}$ $(0  \le \theta_{\text{cp}} < 2\pi)$,
where $\theta_{\text{cp}}$ is called the checkpoint phase.
The $k$th $m$-cycles
period and its variance are defined as
\begin{equation}
T_{m}^{(k)}=t^{(k)}-t^{(k-m)},
\end{equation}
\begin{eqnarray}
V_{m} &= & E[( T_{m}^{(k)} - m \tau)^{2}],
\label{eq:defV}
\end{eqnarray}
where $E[\cdots]$ denotes the statistical average over $k$ and $\tau$ is the average period given by $\tau=E[T_{1}^{{(}k{)}}]$.
Note that in this paper $E[\cdots]$ denotes both the statistical average over $k$ and the ensemble average, which are identical in the steady state.
Note further that  $V_{m}$ is calculated from the spike-time data of one oscillator.
To quantify the relationship between two oscillators,
the standard deviation
of time lag between the spikes of the oscillators is defined as
\begin{equation}
\zeta=\sqrt{E[(t^{(k)}_{1}-t^{(k)}_{2})^{2}]},
\end{equation}
where $t^{(k)}_{i}$ is the $k$th spike time of the $i$th oscillator.

To derive an inference theory, we consider a pair of coupled phase oscillators subject to noise.
When limit-cycle oscillators are weakly coupled to each other and subject to weak noise, the dynamics can be described by~\cite{winfree67, kuramoto84}:
\begin{eqnarray}
\begin{array}{l}
\dot{{\theta_{1}}}
= \omega + \kappa J(\theta_{1},\theta_{2})+  Z(\theta_{1}) \sqrt{D} \xi_{1}(t), \\
\dot{{\theta_{2}}}
 = \omega + \kappa J(\theta_{2},\theta_{1})+     Z(\theta_{2})  \sqrt{D}  \xi_{2} (t) ,
\end{array}
\label{eq:coupled}
\end{eqnarray}
where $\theta_{i}$ is the phase of oscillator $i$ and $\kappa \ge 0$ is the coupling strength.
The independent and identically distributed (i.i.d.) noise $\xi_{i} (t)$ satisfies $E[\xi_{i}(t)]=0$ and $E[\xi_{i}(t) \xi_{j}(t')]=\delta_{ij}\delta(t-t')$.
The positive constant $D$ represents the noise intensity.
The phase sensitivity function $Z(\theta)$ is a $2\pi$-periodic function that quantifies the phase response to noise.
The $2\pi$-periodic function $J(x,y)$ describes the interaction between oscillators that leads to synchronization.
We assume that $J(\theta,\theta)=0$, which is satisfied in systems with diffusive coupling between chemical oscillators or gap-junction coupling between cells.
We focus on systems that are well synchronized in phase.

Our inference methods are based on the formula of period variability.
In previous work~\cite{SDrapid}, the following expression for the variance $V_{1}$ was derived from the system in Eq.~(\ref{eq:coupled}) by means of linear approximation:
\begin{equation}
V_{1} (\theta_{\text{cp}})
= C_{1}+  C_{2} \frac{ {d(\theta_{\text{cp}})}^{2} }{ \omega^{2} },
\label{eq:sigma}
\end{equation}
where $C_1$ and $C_2$ are independent of $\theta_{\text{cp}}$ and given by
$C_{1}=\frac{D}{2} \int_{0}^{2 \pi} \frac{Z(\theta)^{2}}{\omega^{3}}  d\theta$
and
$C_{2}=(1-\exp[c \kappa  ])/2$.
The negative constant $c\kappa$ corresponds to the average effective attractive force between the oscillators over one oscillation period.
That is, $c =  \frac{1}{\omega} \int_{0}^{2 \pi} f_{Y}(\theta) d \theta$, where
$f_{Y} (\theta) \equiv  \left.  \frac{\partial J}{\partial x}\right|_{x=y=\theta} - \left. \frac{\partial J}{\partial y}\right|_{x=y=\theta } $.
The $2\pi$-periodic function
$d(\theta_{\text{cp}}) \equiv \sqrt{ E\left[ \| \theta_{1} - \theta_{2} \|^{2} \right] _{\theta_{1} = \theta_{\text{cp}}}}$
represents the phase distance from in-phase synchronization, where $\| \theta_{1} - \theta_{2} \|$ is the phase difference defined on the ring $[-\pi ,\pi)$.
If $x_k$ is the value of $x(t)$ when $\theta_{1}$ first passes through $2\pi k + \theta_{\text{cp}}$, then $E[x(t)]_{\theta_{1}= \theta_{\text{cp}}}$
represents the average of $x_k$ over $k$.
Note that $d(\theta_{\text{cp}})$ is proportional to $\sqrt{D}$ and dependent on $\kappa$~\cite{SDrapid}.

Through a derivation similar to that of Eq.~(\ref{eq:sigma}), we derive that $V_{m}$ is given by
\begin{equation}
V_{m}(\theta_{\text{cp}})
= m aD +   \frac{[1-\exp( m c \kappa  )]}{2}  \left[  \frac{d(\theta_{\text{cp}})}{\omega}  \right] ^{2} ,
\label{eq:Vm}
\end{equation}
where
$a \equiv \frac{1}{2} \int_{0}^{2 \pi} \frac{Z(\theta)^{2}}{\omega^{3}}  d \theta > 0$. 
See Appendix A 
for the derivation.  
%
Since $a$ represents an average phase response to noise, the product $aD$ represents the effective noise intensity~\cite{kuramoto84}.
Our purpose is now to infer $aD$ and $|c|\kappa$,
which are important values because they determine the strength of the phase diffusion and the time scale of the synchronization, respectively \cite{kuramoto84}.

{\it{Method I.}}
We use only $V_{1}$, $V_{2}$, and $V_{3}$ for one of the oscillators.
Combining Eq.~(6) for $m=1,2,3$, we can determine the three unknowns $aD$,  $c \kappa$, and
$ ( {d}/{\omega}  ) ^{2}$.
In particular, we obtain
\begin{equation}
aD=\frac{-V_{1}^{2}-V_{2}^{2}+V_{1}V_{2}+V_{1}V_{3}}{3(V_{1}-V_{2}  {)} +V_{3}}
\label{eq:aD1}
\end{equation}
and
\begin{equation}
|c|\kappa=  \log{   \frac{ V_{2}- 2V_{1}}{V_{3}-2V_{2}+V_{1}}}.
\label{eq:ck1}
\end{equation}
%
%
%
Note that,
as shown below,
Eq.~(\ref{eq:ck1}) states
that a temporal correlation
 decays exponentially
 with spike times
 and the decay constant is given by the effective coupling strength $|c|\kappa$.
We define the temporal correlation as
\begin{eqnarray}
G_{m}  & = & \frac{1}{n} \sum_{k=1}^{n} [   (T_{1}^{(k-m)} - \tau) (  T_{1}^{(k)} -\tau   ) ].
\end{eqnarray}
Recall that $V_{m}= \frac{1}{n} \sum_{k=1}^{n}  (    \sum_{i=1}^{m} T_{1}^{(k-i+1)}       - m \tau )^{2} $.
When $n$ is sufficiently large, i.e., $n \gg |m|$ and $n \gg | T_{1}^{(k)}  -\tau |^{2}/ V_{m} $ for any $k$,
we obtain
\begin{eqnarray}
 G_{m}   =     \frac{1}{2}  (       V_{m+1}     -2 V_{m}  + V_{m-1}     ),
 \label{eq:c}
  \end{eqnarray}
  where $m \ge 1$.
Thus,
the numerator and  the  denominator in Eq.~(\ref{eq:ck1})
represent the correlations  $G_{1}$ and $G_{2}$, respectively,
i.e., $\exp{(|c| \kappa)}=G_{1}/G_{2}$.


{\it{Method II.}}
We additionally use $\zeta$, which is the standard deviation of the spike-time lags.
%
%
When $D$ and $\kappa$ are sufficiently small, 
we can assume that 
$\dot{\theta}_{i}= \omega +\mathcal{O}(D, \kappa)$. 
%
Using this approximation, we can express $V_{m}$  as
\begin{equation}
V_{m} = m aD +  \frac{\zeta^{2}}{2}  [1-\exp( m c \kappa  )],
\label{eq:Vm2}
\end{equation}
where  $\mathcal{O}(D,\kappa)  \zeta^{2}$ is neglected.
In terms of $V_{1}$, $V_{2}$, and $\zeta$, the two unknowns $aD$ and $c \kappa$ are given by
\begin{equation}
aD=V_{1}-\sqrt{\frac{\zeta^{2}}{2}(2 V_{1}-V_{2}) }
\label{eq:aD2}
\end{equation}
and
\begin{equation}
|c| \kappa=   - \log \{ 1- \sqrt{\frac{2}{\zeta^{2}}(2V_{1}-V_{2})}\}.
\label{eq:ck2}
\end{equation}
Our formulae in Eqs.~(\ref{eq:aD1}) , (\ref{eq:ck1}), (\ref{eq:aD2}), and (\ref{eq:ck2}) are independent of the checkpoint phase, while $V_{m}$ and $\zeta$ are not~\cite{SDrapid}.

We demonstrated the validity of the inference methods with numerical experiments.
First, we again employed the phase oscillator model in Eq.~(\ref{eq:coupled}).
We assumed $J(x,y)=z(x)[h(x)-h(y)]$, which represents gap-junction coupling or diffusive coupling \cite{winfree67,kuramoto84}.
We set $z(\theta)=\sin{\theta}$ for $0 \le \theta < \pi$ and $z(\theta)=0$ for $\pi \le \theta < 2\pi$, with $h(\theta)=\cos{\theta}$.
The region
satisfying
$z(\theta)=0$ mimics the refractory stage that exists for many chemical 
and biological  oscillators. 
We set $Z(\theta)=1$ and $\omega=2\pi$.
Under these assumptions, $a=\frac{1}{2} \frac{1}{(2\pi)^{2}}$,   $|c|=\frac{1}{2}$,  and $\tau=1$.
For $\xi_{1}(t)$ and $\xi_{2}(t)$ we assumed white Gaussian noise.

We prepared 16 parameter sets, each with a different combination of coupling strength and noise intensity given by $\kappa=0.25 \cdot 2 \pi  n_\kappa$ and $D=0.002 \cdot (2\pi)^{2} n_D$, where $n_\kappa,n_D=1,2,3,4$.
We integrated Eq.~(\ref{eq:coupled}) using the Euler scheme with a time step of $5 \times 10^{-4}$.
The initial conditions were $\theta_{1}(0)=\theta_{2}(0)=0$.
In this simulation, we fixed the checkpoint phase at $\theta_{\text{cp}}=\pi/2$ and observed the spike timing for $10^{2}\leq t \leq 10^{6}$.
Three realizations were simulated for each parameter set.
By using the $V_{m}$ of one oscillator, we obtained three pairs of inferred parameters.
By using the $V_{m}$ of the other oscillator, we obtained three additional pairs.
Thus, we have six pairs of inferred values for each parameter set.

\begin{figure}[h]
\begin{center}
\includegraphics[width=5cm]{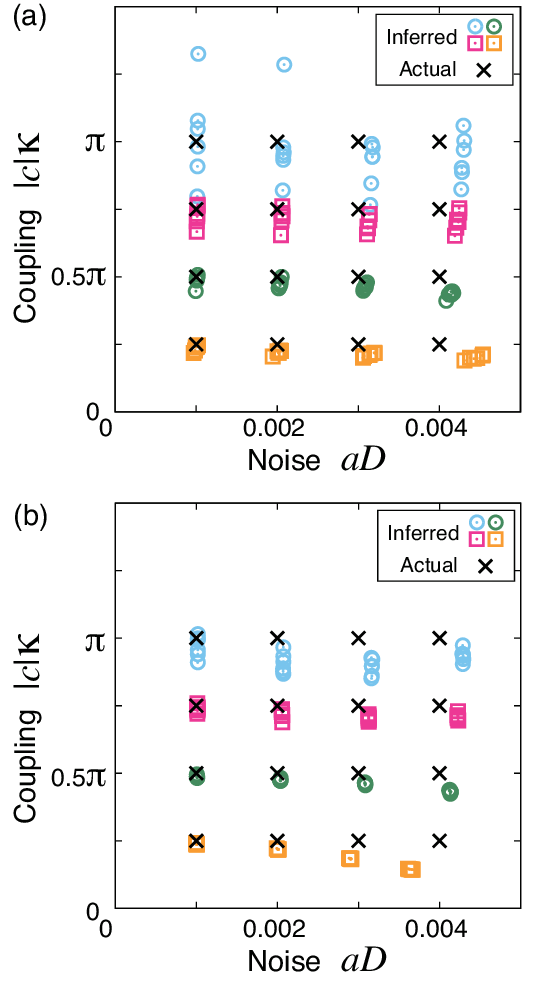}
\end{center}
\caption{(color online)
Simultaneous inferences of effective noise intensity $aD$ and effective coupling strength $|c|\kappa$ obtained with (a) method~I  and (b) method~II.
Actual values are plotted as crosses.
Inferred values are plotted as squares (for $|c| \kappa= 0.25\pi, 0.75\pi$) and
circles
(otherwise).}
\label{fig:phINF}
\end{figure}

\begin{figure}[h]
\begin{center}
\includegraphics[width=8cm]{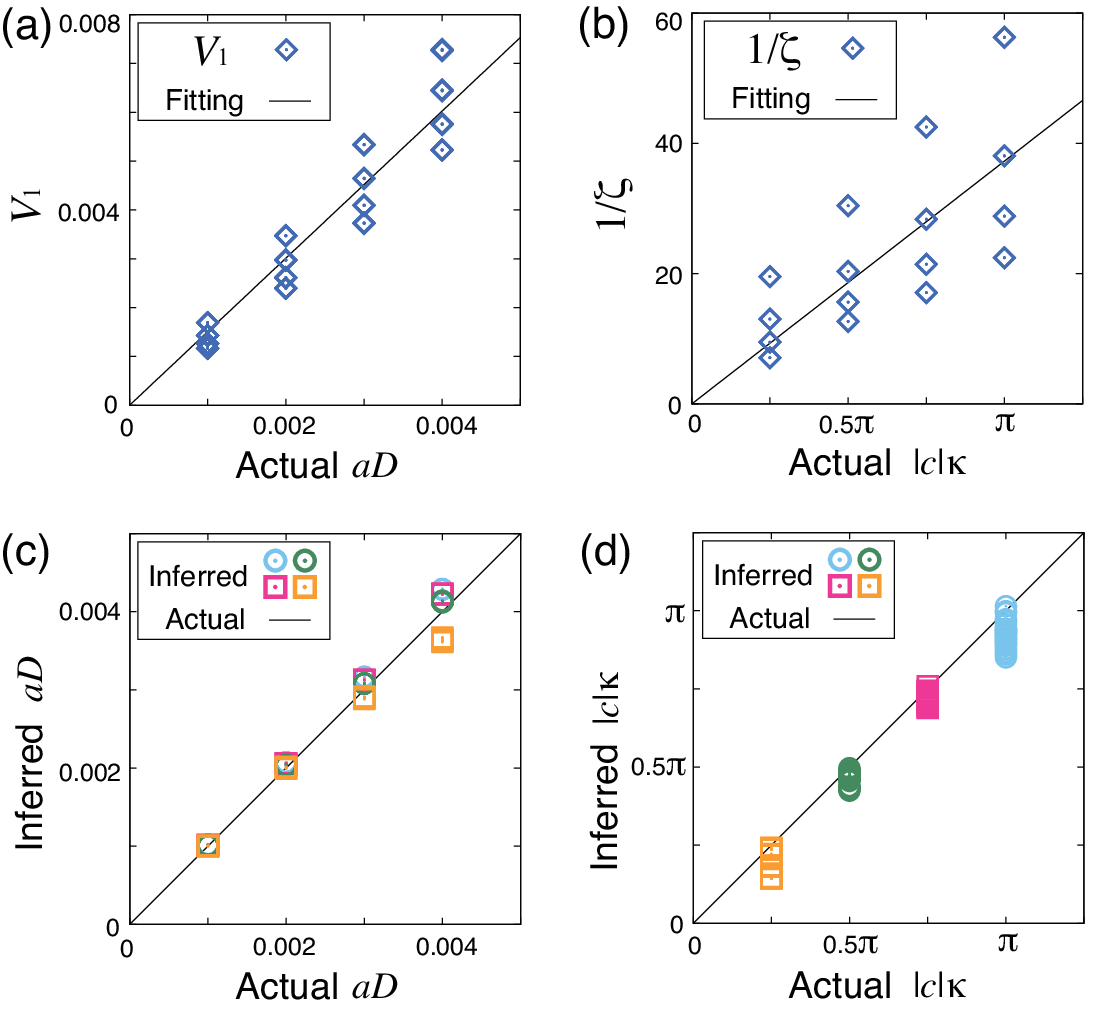}
\end{center}
\caption{(color online).
Raw data on (a) period variance $V_{1}$ versus effective noise intensity $aD$ and (b) inverse time lag ${1/\zeta}$ versus effective  coupling strength $|c| \kappa$.
The lines in (a) and (b)
are drawn by using the least-square method.
For comparison, the inferred values of (c) $aD$ and (d) $|c| \kappa$ with method II
are plotted
versus the actual values of $aD$ and $|c| \kappa$, respectively.
While our methods achieved precise inferences,
 $V_{1}$ and $1/\zeta$ did not.
}
\label{fig:VandZ}
\end{figure}

The results of the simultaneous inferences of noise intensity and coupling strength with methods~I and~II are shown in Figs.~\ref{fig:phINF}(a) and~\ref{fig:phINF}(b), respectively.
In Fig.~\ref{fig:phINF}(a), the inferred values approximately reproduce the actual values even though only one oscillator was observed.
The error in the inference increases as the coupling strength is increased.
In Fig.~\ref{fig:phINF}(b), the inferences by method~II are obviously an improvement on the results of method~I.


We emphasize that a naive use of the statistical values $V_{m}$ and $\zeta$ will not yield successful inferences of
noise intensity and coupling strength.
The correlation between $V_{1}$ and $aD$ is shown in Fig.~\ref{fig:VandZ}(a), and that between $1/\zeta$ and $|c| \kappa$ is shown in Fig.~\ref{fig:VandZ}(b).
We found that
their correlation coefficients
were $0.96$ and $0.70$, respectively.
In contrast,
the correlation coefficient
between the actual and inferred noise intensities (coupling strengths) for method II
was $0.99$ (0.99), as shown in Fig.~\ref{fig:VandZ}(c) ((d)).
This fact indicates that
our methods
are superior over the naive use of $V_{m}$ and $\zeta$.
In addition,
the naive use provides only relative intensities,
whereas our methods
 directly infer the absolute values of $aD$ and $|c| \kappa$.


\begin{figure}[h]
\begin{center}
\includegraphics[width=5cm]{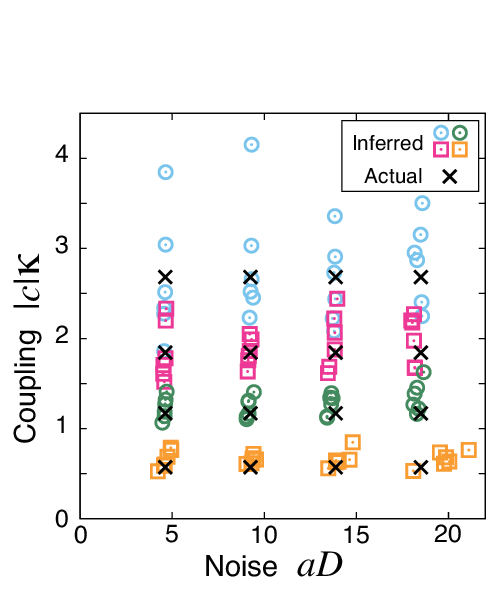}
\end{center}
\caption{(color online).
Simultaneous inference of effective noise intensity $aD$ and effective coupling strength $|c| \kappa$ 
for the FitzHugh-Nagumo model with method~I.
Actual and inferred values are plotted in the same manner as Fig.~\ref{fig:fig1}(b).
The inferences were successful overall,
although the errors became larger as the coupling strength was increased.}
\label{fig:fnINF}
\end{figure}

Next, we demonstrated the inference method for a more realistic system.
Specifically, we employed the 
{coupled}
FitzHugh-Nagumo 
{oscillators,}
described by
\begin{eqnarray}
\begin{array}{l}
\dot{v_{{i}}}= v_{{i}}(v_{{i}}- \!\alpha)(1-\!v_{{i}})-\!w_{{i}} + \!\kappa (v_{{j}}-\!v_{{i}}) + \!\sqrt{D} \xi_{{i}}(t),     \ \ \ \\
\dot{w_{{i}}} = \epsilon (v_{{i}}- \beta w_{{i}}), 
\end{array}
\label{eq:FN}
\end{eqnarray}
{for $(i,j)=(1,2)$ and $(2,1)$.}
We set $\alpha=-0.1$, $\beta=0.5$, and $\epsilon=0.01$.
Each $\xi_{i}(t)$ was the same as that in the inferences discussed above.
This system shows limit-cycle oscillation with a period $\tau \simeq 126.5$ when noise and coupling are absent.

The actual values of $a$ and $c$
for this system were obtained as follows:
to calculate $a$, we numerically integrated the function $Z$ representing the phase response to noise; and
to calculate $c$, we observed the relaxation of the phase difference between two oscillators in a system with a fixed $\kappa$ but without noise.
The phase difference
is expected to exponentially decrease by a factor of $\exp(c \kappa)$ each period.
We adopted the value of $|c| \kappa$ obtained from this relaxation as the effective coupling strength.
%

We prepared $16$ parameter sets with $D=10^{-4}n_D$ and $\kappa=10^{-2}n_\kappa$, where $n_D, n_\kappa = 1,2,3,4$.
We integrated Eq.~(\ref{eq:FN}) using the Euler scheme with a time step of $ 10^{-3}$.
In this simulation, the checkpoint threshold was fixed at $v_{\text{cp}}=0.6$ and the $k$th spike time $t^{(k)}$ was defined as the time at which $v$
first
passes through  $v_{\text{cp}}$ in the $k$th oscillation.
We observed the spike timing for $2 \times 10^{3} \leq t \leq  2 \times 10^{7}$.
The observed number of oscillations was about $1.6\times 10^5$, corresponding to a day in the experiment on cardiac myocytes~\cite{yamauchi02}.
Three realizations were simulated for each parameter set.

The results of simultaneous inferences for the FitzHugh-Nagumo model with methods~I and~II are shown in Figs.~\ref{fig:fnINF} and \ref{fig:fig1}(b), respectively.
Figure~\ref{fig:fnINF} reveals that precise inferences were obtained with method~I , except in the case of the greatest coupling strength, even though data on only one oscillator were employed.
Figure~\ref{fig:fig1}(b) reveals that better inferences were obtained with method~II , because the additional information raised the precision of each inference.

The causes of the inference error in the numerical demonstrations are summarized as follows:
(i) Equations~(\ref{eq:Vm}) and~(\ref{eq:Vm2}) were derived using linear approximation.
As the noise is increased, these equations deviate from reality.
(ii) The $V_{m}$ and $\zeta$ obtained numerically are different from the actual values because the observation time is finite.
(iii) There is a limit to the precision in the determination of spike timing.
In our demonstrations, this limit corresponds to the time step in the numerical integration.
(iv) The FitzHugh-Nagumo model can become poorly approximated by the phase model as noise intensity or coupling strength is increased.
In Appendix B, 
we demonstrate that the error caused by (ii)  can be estimated via the  bootstrap method without repeating the inference experiments. 

The values inferred must be influenced by a complex combination of these error causes.
We discuss here the fact that
the largest errors were found in the inferences for high coupling strengths.
For large $|c| \kappa$,
the terms of  $e^{3 c \kappa}  d^{2} $ and $e^{2 c \kappa} \zeta^{2} $
are
 the smallest in Eqs.~(\ref{eq:Vm}) and (\ref{eq:Vm2}), respectively.
The magnitude of these terms
 may be comparable to the errors mentioned above
when coupling is sufficiently strong,
because of which our inference methods become imprecise.
%
One of the reasons that Method II achieved a more precise inference is that
it avoids using $ e^{3 c \kappa} d^{2}$.

Finally, we discuss the applicability of our methods.
In the abovementioned experiment on cardiac cells~\cite{yamauchi02},
the period variance was gradually decreased over about ten days of the culture.
By assuming that this evolution is sufficiently slow, we can infer the coupling strength and noise intensity for each day with our methods.
Hence, it is possible to estimate the growth process of a cell culture noninvasively.

This paper has proposed theoretical methods to infer simultaneously the coupling strength and noise intensity from spike-time data alone.
Although statistics including $V_{m}$ and $\zeta$ are dependent on both parameters, our method can distinguish between the effects of noise and coupling without an external perturbation.
{
In Appendix C, 
we confirm that our theory provides 
the accurate relative relationship between inferred parameters even when two oscillators have slightly different frequencies.
Moreover, 
as briefly explained 
in Appendix A, 
inference formulae for a population of oscillators with all-to-all coupling 
can be derived 
in the same manner as the case for $N=2$.}
Therefore,
our methods are ready for applications to real systems.

\begin{acknowledgments}
This work was supported by {JSPS KAKENHI Grant Number JP19K03663, JP21K12056} 
and 23$\cdot$11148.  
\end{acknowledgments}


\appendix
\renewcommand{\thefigure}{A\arabic{figure}}
\setcounter{figure}{0}

\section{Derivation of Eq.~(6)}

The proposed method is based on the fact that 
the period variability is described in terms of coupling strength and noise intensity,  
as shown in a previous study [21]. 
Because Ref.~[21] has dealt with only $V_{1}$, 
 we  present the derivation of $V_{m}$ $(m=1,2,3)$ in Eq.~(6). 
Note that $J(\theta, \theta)=0$ is assumed  while it is not in Ref. [21]. 
 The derivation consists of
two steps:  
(i) calculation of
the phase diffusion $\sigma_{m} (\theta_{\text{cp}})^2$ [defined by
Eq.~(\ref{eq:sigma-m})]
with a linear approximation, and (ii) transformation from 
$\sigma_{m} (\theta_{\text{cp}})^2$ to $V_{m}(\theta_{\text{cp}}$)
[defined by Eq.~(\ref{eq:trans})].

First, 
let us present a few  notations  and  assumptions. 
In the absence of noise ($D=0$) in the coupled-oscillator model Eq.~(4),  the oscillators are assumed to 
synchronize in phase, i.e., $\theta_{1,2}(t) \rightarrow \phi (t) $
$(t\rightarrow \infty) $, where $\phi (t)$  is a solution of
\begin{equation}
\dot{\phi} (t)=\omega+ \kappa J(\phi,\phi) = \omega, 
\label{eq:phi}
\end{equation}
{because of the assumption $ J(\phi,\phi) =0$.}
{Thus,  $\phi(t) = \omega t + \text{C}$, and we choose $\text{C}=0$ without loss of generality.}
The necessary condition for the stability of in-phase synchrony for
$D=0$ is provided below [see Eq.~(\ref{eq:nc})]. 
We define  the time $t_{\text{cp}}$ by
$\phi(t_{\text{cp}})=\theta_{\text{cp}}$.
Although $\tau$ was introduced as the average period in the main text, 
 it is actually equal to the oscillation period observed for $D=0$ owing to the weak-noise assumption {[as explained around Eq.~(\ref{eq:aveT})]}. 
 Therefore, 
 $\phi( t_{\text{cp}} +\tau)= \theta_{\text{cp}}+2\pi$. 
%
{In the presence of noise, our system approaches the steady state after a transient time. In the steady state,}
\begin{equation}
P(\|  \theta_{1}-\theta_{2} \|; \theta_{1}=\Psi)=P(\|  \theta_{1}-\theta_{2} \|  ; \theta_{1}=\Psi+2\pi),
\label{eq:steady}
\end{equation}
{is satisfied for all $\Psi$,}
where $P(\|  \theta_{1}-\theta_{2} \|; \theta_{1}=\Psi) $ is the
probability density function of the distance $\|
\theta_{1}-\theta_{2} \|$
at $\theta_{1}= \Psi$.
%
We assume that the system is in the steady state at $t=0$.
The ensemble we consider here is defined by the initial condition at
$t=t_{\text{cp}}$ {$(\ge 0)$}, $\theta_{1}(t_{\text{cp}})=\theta_{\text{cp}}$,
and $\theta_{2}(t_{\text{cp}})$ is distributed 
 in
 $[\theta_{1}(t_{\text{cp}})-\pi, \theta_{1}(t_{\text{cp}}) + \pi)$ according to Eq.~(\ref{eq:steady}), 
 {where the average of $\theta_{2}(t_{\text{cp}})$ is assumed to be $\theta_{\text{cp}}$.} 
Henceforth, 
 $E[\cdots]$ represents the average taken over this
ensemble, 
{e.g., $E[\theta_{2}(t_{\text{cp}}) ] =\theta_{\text{cp}}$.}
The phase diffusion $\sigma_{m} (\theta_{\text{cp}})$ is defined by
\begin{equation}
 \sigma_{m} (\theta_{\text{cp}})^2 =
E[ (\theta_{1}(t_{\text{cp}}+ m \tau) -\theta_{1}(t_{\text{cp}})-2 \pi m)^{2} ], 
\label{eq:sigma-m}
\end{equation}
where  $m= 1, 2,3$.
Because the noise intensity $D$ is sufficiently small, 
and  the other parameters and functions are of $\mathcal{O}(1)$,
the phase difference $\| \theta_1 - \theta_2 \|$ is small in
most cases in the steady state.

To calculate the phase diffusion, we decompose $\theta_{1,2} $ as
$\theta_{1,2} (t) =\phi(t) + \Delta_{1,2}(t)$.
Then, 
 the time duration $0\le t \le \mathcal{O}(\tau)$ is considered, in which
$\Delta_{1,2}(t) \ll 1$ is expected in most cases because $D\ll 1$.
Therefore, we can linearize Eq.~(4). 
We define the two modes, $X=\Delta_{1}+\Delta_{2}$ and
$Y=\Delta_{1}-\Delta_{2}$, which obey
\begin{eqnarray}
\dot{X}(t)&=&\xi_{X} (t,\phi(t)),   \label{eq:Xdot}    \\
\dot{Y} (t)&=&\kappa
f_{Y} (\phi(t)) Y(t)
+
\xi_{Y} (t,\phi(t)),
\label{eq:Ydot}
\end{eqnarray}
where 
$f_{Y} (\phi) \equiv  \left.  \frac{\partial J}{\partial x}
\right|_{x=y=\phi} -   \left.     \frac{\partial J}{\partial y}
\right|_{x=y=\phi } $,
$\xi_{X,Y}(t,\phi(t)) \equiv  \sqrt{D}
Z(\phi(t)) (\xi_{1}(t) \pm \xi_{2}(t))$,
{and we use $ \left.  \frac{\partial J}{\partial x}
\right|_{x=y=\phi} +  \left.     \frac{\partial J}{\partial y}
\right|_{x=y=\phi } = \frac{d J(\phi,\phi)}{d \phi} =0$.}
The solutions 
of Eqs.~(\ref{eq:Xdot}) and (\ref{eq:Ydot})  can be described as
\begin{eqnarray}
  {X}(t) &=&       X(0)
+ \int_{0}^{t}  \xi_{X}(t',\phi(t'))  dt'  ,   \label{eq:xt} \\
  {Y}(t) &=& \exp{ \left[+\kappa F_{Y}(\phi(t)) \right] }    \times     \left\{   Y(0)
+ \int_{0}^{t} \exp{  \left[ - \kappa F_{Y}(\phi(t')) \right] } \xi_{Y}(t',\phi(t'))  dt'                 \right\}  \label{eq:yt},
\end{eqnarray}
where
$F_{Y}(\phi(t)) \equiv \int_{0}^{t}  f_{Y}(\phi(t')) dt'$.
For $\xi_{Y}=0$, we obtain
$F_{Y}(2\pi)=(1/\kappa)\ln(Y(\tau)/Y(0))$.
Therefore, in the absence of noise, in-phase synchronization is
stable if
\begin{equation}
F_{Y}(2 \pi)  \equiv c <0.
\label{eq:nc}
\end{equation}

{
The conditions 
$\phi(t_{\text{cp}})=\theta_{\text{cp}}$, 
$\theta_{1}(t_{\text{cp}}) =\theta_{\text{cp}}$ and $E[\theta_{2}(t_{\text{cp}})] =\nolinebreak \theta_{\text{cp}}$,  
imply 
 $\Delta_{1}(t_{\text{cp}}) =0$ and $E[\Delta_{2}(t_{\text{cp}})]  =0$, which lead to 
   $E[X(t_{\text{cp}})]=0$ and $E[Y(t_{\text{cp}})]=0$.
Combining these conditions for $t=t_{\text{cp}}$ with the ensemble averages of Eqs.~(\ref{eq:xt}) and (\ref{eq:yt}) described as 
\vspace{-0.5cm}
\begin{eqnarray}
  E[{X}(t)] &=&     E[  X(0) ],   \label{eq:Ext} \\
E[  {Y}(t)] &=& \exp{ \left[+\kappa F_{Y}(\phi(t)) \right] }    \times    E[   Y(0)]     \label{eq:Eyt}, 
\end{eqnarray}
we obtain 
$E[X(0)]=0$ and  $E[Y(0)]=0$.
Consequently, 
$E[X(t)]=E[Y(t)]=0$ holds for any $t$. 
Therefore, we find that the phase $\theta_{1}$ is  increased by $2 \pi$ on average during $\tau$  as follows:  
\begin{eqnarray}
&&E[ \theta_{1} (t_{\text{cp}}+ \tau ) - \theta_{1} (t_{\text{cp}}) - 2 \pi ]   \non \\
&=& E[     \Delta_{1} (t_{\text{cp}}+\tau) -           \Delta_{1} (t_{\text{cp}})]             = E[\Delta_{1} (t_{\text{cp}}+\tau)]  \non \\
 &= & \frac{1}{2} \Bigg( E[ X (t_{\text{cp}}+\tau)] +E[Y(t_{\text{cp}}+\tau)]   \Bigg) =0. 
 \label{eq:aveT}
\end{eqnarray}
This indicates 
that  the average period $E[T_{1}^{(k)}]$ is equal to the period $\tau$ observed for $D=0$ in the linear approximation theory.}


The correlations of noise terms, 
$\xi_{X,Y}(t,\phi(t)) =  \sqrt{D}Z(\phi(t)) (\xi_{1}(t) \pm \xi_{2}(t))$,  
are given as
\begin{eqnarray}
E[ \xi_{X}(s,\phi(s))  \xi_{X}(s',\phi(s'))]  
  & =  & 2 D Z(\phi(s)) Z(\phi(s'))  \delta(s-s'),   \label{eq:noixx}  \\
  E[ \xi_{Y}(s,\phi(s))  \xi_{Y}(s',\phi(s'))]  
  & =  & 2 D Z(\phi(s)) Z(\phi(s'))  \delta(s-s'),     \label{eq:noiyy} \\
    E[ \xi_{X}(s,\phi(s))  \xi_{Y}(s',\phi(s'))]  
  & =  & 0.   \label{eq:noixy} 
\end{eqnarray}
%
  Using  Eqs.~(\ref{eq:xt}), (\ref{eq:yt}),  (\ref{eq:noixx}), (\ref{eq:noiyy}), (\ref{eq:noixy}), and $E[\xi_{X,Y}(t,\phi(t))]=0$, 
we obtain 
\begin{eqnarray}
 E[X(t)^{2}]  &=&  
E[X(0)^{2}]    +  \int_{0}^{t}  \int_{0}^{t}     E[ \xi_{X}(s,\phi(s))  \xi_{X}(s',\phi(s'))]     ds ds'
   \nonumber  \\
& =& 
E[X(0)^{2}] 
 +2D \int_{0}^{t}  Z(\phi(s))^{2}     ds ,  \label{eq:xx}  \\
 E[Y(t)^{2}]  &=&  \exp{[ 2 \kappa F_{Y} (\phi(t))   ]}  
  \biggl[
E[Y(0)^{2}]   \nonumber   \\ 
&& +  \int_{0}^{t}  \int_{0}^{t}   \exp[-\kappa \{F_{Y}(\phi (s))+ F_{Y}(\phi (s'))\}]   E[ \xi_{Y}(s,\phi(s))  \xi_{Y}(s',\phi(s'))]     ds ds'
\biggr]       \nonumber  \\
& =& \exp{[ 2 \kappa F_{Y} (\phi(t))   ]} 
  \biggl[
E[Y(0)^{2}] 
 +2D \int_{0}^{t}  Z(\phi(s))^{2}      \exp[-2\kappa F_{Y}(\phi (s))] ds  \biggr],   
\end{eqnarray}
and 
\begin{eqnarray}
E[X(t)Y(t)] &  = & \exp{[  \kappa   F_{Y} (\phi(t))   ]}  
E[X(0)Y(0)]. \label{eq:xy}
\end{eqnarray}
Putting $t=t_{\text{cp}}+\tau$ in 
 Eqs.~(\ref{eq:xx}) and (\ref{eq:xy}), 
we obtain

\begin{eqnarray}
E[X(t_{\text{cp}} +\tau)^{2}] 
&=& 
   E[X(0)^{2}] 
 +2D \int_{0}^{t_{\text{cp}}+\tau}  Z(\phi(s))^{2}       ds  \nonumber \\
 &=& 
   E[X(0)^{2}] 
   +2D \int_{0}^{t_{\text{cp}}}  + \int_{t_{\text{cp}}}^{t_{\text{cp}}+\tau}
     Z(\phi(s))^{2}      ds     \nonumber \\
     & = &
     E[X(t_{\text{cp}} )^{2}]
     + 2D 
\int_{0}^{\tau}
Z(\phi(s))^{2}  ds,  
\label{eq:xxtau}
\end{eqnarray}
and 
\begin{eqnarray}
E[X(t_{\text{cp}} +\tau) Y(t_{\text{cp}} +\tau)] 
&=& \exp{[  \kappa  F_{Y} ( \theta_{\text{cp}}+2 \pi )  ]}   E[X(0)Y(0)]   \nonumber \\
&  = &  \exp{[ \kappa c  ]}  E[X(t_{\text{cp}} )  Y(t_{\text{cp}} )  ],     
\label{eq:xytau}
\end{eqnarray}
where we use $\phi(t_{\text{cp}}+\tau)=\theta_{\text{cp}}+2 \pi$, 
$F_{Y}(\theta+2 \pi)=F_{Y}(\theta) +c$, and  
$Z(\theta+2 \pi)=Z(\theta)$. 
{Because $\Delta_{1}(t_{\text{cp}})=0$,}   
we have 
\begin{equation}
E[X(t_{\text{cp}})^{2}]
=-E[X(t_{\text{cp}}) Y(t_{\text{cp}}) ] =E[Y(t_{\text{cp}})^{2}]
=d(\theta_{\text{cp}})^{2}. 
\label{eq:d2}
\end{equation}

Eq.~(\ref{eq:steady}) can be rewritten as 
$P(| \Delta_{1}-\Delta_{2} |; t ) \cong P( |  \Delta_{1}-\Delta_{2} |  ;
t+\tau)$; 
then $ E[Y(t)^{2}]= E[Y(t+\tau)^{2}]$ approximately holds true,
leading to
\begin{equation}
E[Y(0)^{2}]=2D \frac{\exp[2 \kappa c ]} { 1-\exp[2 \kappa c ]}
 \int_{0}^{\tau}   Z(\phi(t'))^{2} \exp[-2 \kappa F_{Y}(\phi(t'))] dt'.
\end{equation}
%
In addition, because $d(\theta_{\text{cp}})^{2} = E[
(\Delta_{1}(t_{\text{cp}}) - \Delta_{2}(t_{\text{cp}}))^{2}] =
E[Y(t_{\text{cp}})^{2}]$, we obtain
\begin{eqnarray}
&&{d(\theta_{\text{cp}})}^{2}    = \exp[2\kappa F_{Y}(\theta_{\text{cp}})]    
\left(  E[Y(0)^{2}]  +  2D \! \int_{0}^{\theta_{\text{cp}}}    Z(\phi)^{2}
  \exp[-2\kappa F_{Y}(\phi)] \frac{1}{\omega}  d \phi  \right),
\label{eq:dsyn}
\end{eqnarray}
which is generally $\theta_{\text{cp}}$-dependent even if $Z(\phi)$ is constant.

Using $\Delta_{1}(t_{\text{cp}})=0$, 
Eqs.~(\ref{eq:xxtau}), (\ref{eq:xytau}), (\ref{eq:d2}), and the relation $E[Y(t_{\text{cp}} +\tau)^{2}]=E[Y(t_{\text{cp}} )^{2}]$, 
the following expression is obtained 
for
the phase diffusion: 
\begin{eqnarray}
 \sigma_{m} (\theta_{\text{cp}})^2
   & = & E[(\Delta_{1}(t_{\text{cp}}+m \tau) -\Delta_{1}(t_{\text{cp}}) )^{2}]     \nonumber \\
&=& \frac{1}{4} \left\{    E[X(t_{\text{cp}}+m \tau)^{2}] +E[Y(t_{\text{cp}}+m \tau)^{2}] +2E[X(t_{\text{cp}}+m \tau) Y(t_{\text{cp}}+m \tau)]       \right\}      \nonumber   \\
&=& \frac{1}{4} \left\{  m  \times 2D 
\int_{0}^{\tau}
Z(\phi(t'))^{2}  dt' 
+2(1- \exp{[ m \kappa c  ]} ) {d}(\theta_{\text{cp}})^{2}
    \right\}      \nonumber   \\
    &=&    m  \times \frac{D}{2} 
\int_{0}^{\tau}
Z(\phi(t'))^{2}  dt' 
+\frac{1- \exp{[ m \kappa c  ]} }{2} {d}(\theta_{\text{cp}})^{2}
          \nonumber   \\
          & =& m a D \omega^{2} 
+ \frac{1 -\exp[m c \kappa]}{2}
 {d}(\theta_{\text{cp}})^{2}. 
\label{eq:pdif}
\end{eqnarray}


When the noise intensity is low, most of
the trajectories of $\theta_{1}(t)$ are 
in close proximity 
to the
unperturbed trajectory $\phi(t)$.  
 In such a case, the following relation approximately
holds true:
\begin{equation}
\sqrt{
\frac{ \sigma_{m} (\theta_{\text{cp}})^2}
{V_{m}(\theta_{\text{cp}})} 
}
= \dot{\phi}(\theta_{\text{cp}})  \ \ (= \omega).
\label{eq:trans}
\end{equation}
The same approximation 
was employed
and numerically verified 
 in
Refs.  [21,22].
Using the transformation 
from 
$\sigma_{m} (\theta_{\text{cp}})$ to
$\sqrt{V_{m}(\theta_{\text{cp}})}$,  
we finally obtain Eq.~(6). 

{
We have discussed solely paired identical phase oscillators;  however, 
our theory can easily be extended to 
$N$ globally coupled (all-to-all) identical phase oscillators. 
In fact, the inference formulae for the $N$-oscillator system
are same as those obtained for $N=2$ as shown below.
 When $\theta_{i} $ ($i=1, \cdots, N$)  are decomposed as 
$\theta_{i} (t) =\phi(t) + \Delta_{i}(t)$, 
 the $N$ modes, 
 $X(t)$ and $Y_{j}(t)$ ($j=2, \cdots N$),  are defined as 
\begin{eqnarray}
X(t) &=& \Delta_{1} (t) +  \Delta_{2} (t) + \cdots +  \Delta_{N} (t),       \\
Y_{j} (t) &=& \Delta_{1} (t)   -  \Delta_{j} (t), 
\end{eqnarray}
which obey 
\begin{eqnarray}
\dot{X}(t)&=&\xi_{X} (t,\phi(t)),   \label{eq:XdotALL}    \\
\dot{Y_{j}} (t)&=&\kappa
f_{Y} (\phi(t)) Y_{j}(t)
+
\xi_{Y_{j}} (t,\phi(t)),
\label{eq:YdotALL}
\end{eqnarray}
where 
$f_{Y} (\phi) \equiv  (N-1)   \left.  \frac{\partial J}{\partial x}
\right|_{x=y=\phi} -   \left.     \frac{\partial J}{\partial y}
\right|_{x=y=\phi } $,
$\xi_{X}(t,\phi(t)) \equiv  \sqrt{D} Z(\phi(t)) (\xi_{1}(t)      +    \cdots + \xi_{N}(t))$, 
and 
$\xi_{Y_{j}}(t,\phi(t)) \equiv  \sqrt{D} Z(\phi(t)) (\xi_{1}(t) - \xi_{j}(t))$.
In the same manner as the case for $N=2$, 
\begin{eqnarray}
V_{m} (\theta_{\text{cp}}) &=& 
  m \frac{1}{N}D   \frac{1}{\omega^2} \int_{0}^{\tau} Z(\phi(s))^{2}  ds    \non \\
   &&
+2 \frac{N-1}{N^2} (1-  \exp{[ m \kappa c  ]} ) \frac{1}{\omega^2} \Big(     E[ \Delta_j (t_{\text{cp}} ) ^2   ]  +(N-2) E[ \Delta_j (t_{\text{cp}} )  \Delta_{j'(\neq j)} (t_{\text{cp}} )  ]      \Big),  \non \\
\end{eqnarray}
is obtained. 
These equations for period variabilities ($m=1,2,3$) provide  
 the same inference formulae as Eqs.~(7) and (8) in the main text  
 except for the definitions of $a$ and $c$: 
\begin{eqnarray}
a &=& \frac{1}{N} \int_0 ^{2 \pi} \frac{Z(\theta)^2}{\omega^3} d \theta,  
\end{eqnarray}
and $c$ is defined using  $f_{Y} (\phi) =  (N-1)   \left.  \frac{\partial J}{\partial x}
\right|_{x=y=\phi} -   \left.     \frac{\partial J}{\partial y} \right|_{x=y=\phi } $.} 
{Moreover, the inference formulae Eqs.~(12) and (13) also hold true when we replace $\zeta^2$ with $\hat{\zeta}^2$ defined as}
\begin{eqnarray}
\hat{\zeta}^2 & = & 
4 \frac{N-1}{N^2}  \Big(      E[ (t_1^{(k)} - t_j^{(k)}  )^2 ]  +(N-2) E[ (t_1^{(k)} - t_j^{(k)}  )   (t_1^{(k)} - t_{j'(\neq j)}^{(k)}  )     ]      \Big), 
\end{eqnarray}
because 
\begin{eqnarray}
\frac{E[ \Delta_j (t_{\text{cp}})^2 ]}{\omega^2} & =&  E[ (t_1^{(k)} - t_j^{(k)}  )^2 ],     \\
\frac{E[ \Delta_j (t_{\text{cp}})    \Delta_{j' (\neq j)} (t_{\text{cp}})    ]}{\omega^2} & =&  E[ (t_1^{(k)} - t_j^{(k)}  )   (t_1^{(k)} - t_{j'(\neq j)}^{(k)}  )     ].     \\
\end{eqnarray}


\clearpage

\section{Error estimations through the bootstrap method}

As discussed in the main text, one of the main causes of errors is the finiteness of  observation time. It would be helpful to estimate the magnitude of such errors.
A simple method to estimate the errors is to repeat observations and inferences that help obtain the distribution of the inferred values. 
However,  we intend to avoid repeating the inference experiments 
solely to estimate the errors in the inferred values. 
Now, we  demonstrate that the bootstrap method can estimate the errors in the inferred values.


\begin{figure}[h]
  \begin{center}
         \includegraphics[clip,width=15.0cm]{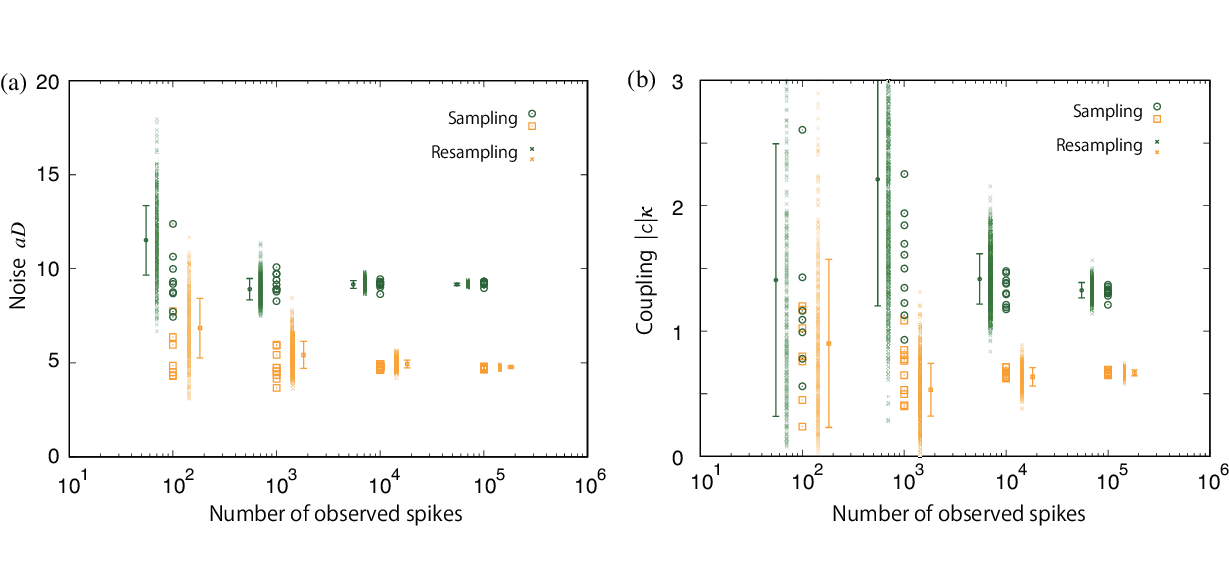}
    \caption{Comparison of the values inferred from the observed spike-time data (open circles and open squares)  
    with those obtained from the bootstrap samples (cross marks).  
  (a) Effective noise intensity and (b)  effective coupling strength in the FitzHugh-Nagumo model in Eq.~(14), 
of  which parameters were set to $D=10^{-4} \times 1$ and  $\kappa=10^{-2} \times 1$ (orange) and   $D=10^{-4} \times 2$ and $\kappa=10^{-2} \times 2$ (green),   
    were estimated  by using the method II.  These are  plotted as a function of the number of observed spikes. 
   The error bars represent standard deviations of the values obtained from the bootstrap samples. 
   As the number of the observed spikes increases,  
   the standard deviations decrease along with 
    the variance of values inferred from the observed data.   
Thus, the standard error in the proposed inference method can be estimated via bootstrap resampling.     }
    \label{fig:bs}
  \end{center}
\end{figure}

First, we adopt the simple method, i.e., we generate the distribution of the inferred values via numerical simulations.
We employed the FitzHugh-Nagumo model  [Eq.~(14)] 
and prepared 
the two parameter sets  ($D=10^{-4} \times 1$, $\kappa=10^{-2} \times 1$ ) [orange  in Fig.~A1] and   ($D=10^{-4} \times 2$, $\kappa=10^{-2} \times 2$) [green]. 
For each parameter set, 
we  simulated ten realizations. 
{For each realization, we  observed $(n+2)$ spikes generated by the oscillators 1 and 2,  where  $n=10^2, 10^3, 10^4, 10^5$.
The spike-timing data $t_{1}^{k}$ and $t_{2}^{k}$   ($k=1, \cdots, n$) yielded the data series 
 $T_{1}^{(k)} = t_{1}^{(k)}-t_{1}^{(k-1)}$, $T_{2}^{(k)} = t_1^{(k)}-t_1^{(k-2)}$, and $ t_{1}^{(k)} - t_2^{(k)}$ 
 ($k=1, \cdots, n$), and noise intensities and coupling strengths were inferred using the method II (squares and circles) in Fig.~A1.}
We can see that the range of the distribution of the inferred values  is sufficiently smaller than  
 the gap between two types of actual values 
when $n=10^4$.  
When $n=10^3$, relative relationships between the two parameter sets can be inferred; however, the  absolute values of the inferred values are imprecise.
Thus, we could confirm that the errors depend on the length of the observation time and determine the number of spikes required for the inference.

To estimate this kind of error without repeating simulations/experiments, 
 the bootstrap method is employed.
 We assumed that only one data set of $n+2$ spikes is available.  A bootstrap sample is given as a data set 
  $(T_{1}^{(k)}, T_{2}^{(k)},  t_{1}^{(k)} - t_2^{(k)})$  for $k=k_1, \ldots , k_n$, where $k_l$ is a random integer in the range  $[1,n]$. Repeating this procedure, we generated 1000 bootstrap samples.
%
Then, we calculated the inferred values from these samples.
The results are shown in Fig.~A1 (cross marks), 
 where the error bars represent the standard deviations of the distributions of the inferred values. 
We see that the error bars obtained from the bootstrap samples can help predict the distribution of the inferred values in the repeated simulations. 
 Therefore, we conclude that 
 the bootstrap method can estimate the errors resulting from the proposed inference method with finite-time observation, 
 and 
 we do not need to repeat 
 the experiments solely for the error estimation.

\newpage
\section{Robustness of the {inference methods against oscillator inhomogeneity}}

In our theory, the two oscillators employed are assumed to be identical.
Here, we consider two oscillators with different
natural frequencies $\omega_1$ and $\omega_2$.
We numerically demonstrate that the proposed inference methods still work accurately
for small values of $\Delta \omega \equiv \omega_1 - \omega_2$.

We modified  the individual frequencies in the phase oscillators given in  Eq.~(4) to
 $\omega_1 = \omega + \frac{\Delta \omega}{2}$ and $\omega_2 = \omega - \frac{\Delta \omega}{2}$. 
The two parameter sets  $(\kappa=0.25 \cdot 2 \pi \times 1, D=0.002 \cdot (2 \pi)^2 \times 1)$ [orange in Fig.~A2] 
and $(\kappa=0.25 \cdot 2 \pi \times 
2, D=0.002 \cdot (2 \pi)^2 \times 2)$  [green] were employed. 
In Fig.~A2, the noise intensities and coupling strength were inferred using the method II.
Two orange squares and two green circles  at  a fixed $ \frac{\Delta \omega}{\omega}$
correspond to the inferred values  for each parameter set 
from the spike-time  data of oscillator 1 and 2. 
  When $ \frac{\Delta \omega}{\omega}=0.09$, 
the value of one of the oscillators could not be obtained because 
 $2V_1 -V_2$ in  
 the inference formulae [Eqs.~(12) and (13)] did not satisfy the condition $2V_1 -V_2  \ge 0$.

Figure A2 indicates that,  
the inferred noise intensities 
(coupling strengths) 
can provide an approximate relative ratio of 1:2  
{when the difference is sufficiently small. 
Furthermore, the relative relationship between the inferred parameters can be accurately obtained 
 while the relative ratio is imprecise when $\frac{\Delta \omega}{\omega}$ is large.
In this sense, our method can be applied to a pair of inequivalent oscillators.}  

\begin{figure}[h]
  \begin{center}
         \includegraphics[clip,width=14.0cm]{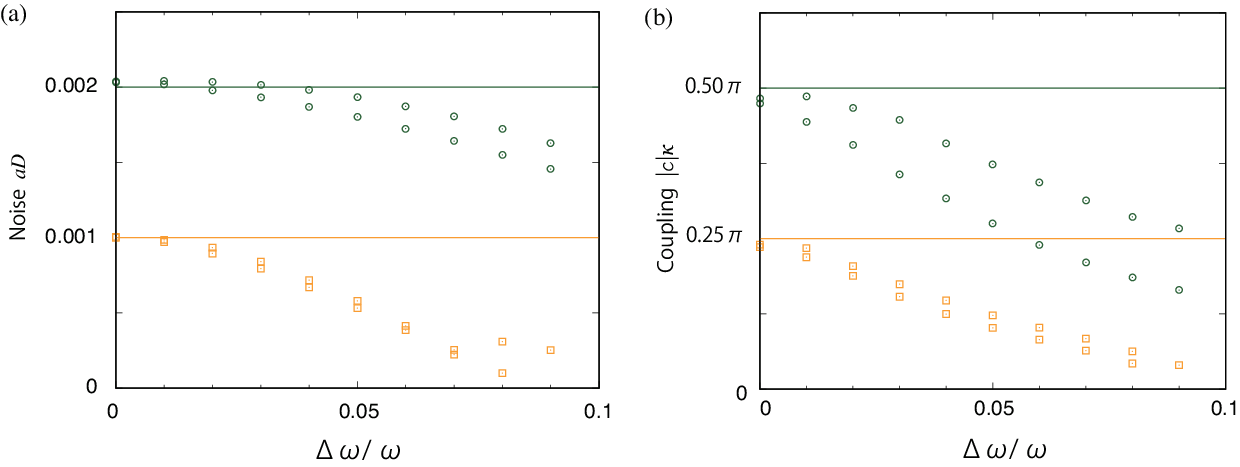}
    \caption{
  (a) Effective noise intensity and (b) effective coupling strength in the phase-oscillator model in Eq.~(4), 
of  which parameters were set to 
$D=0.002 \cdot (2\pi)^2 $ and  $\kappa=0.25 \cdot 2 \pi $ (orange) and  
$D=0.004 \cdot (2\pi)^2 $ and  $\kappa=0.50 \cdot 2 \pi $   (green),   
    were estimated  by using the method II.  These values are  plotted as a function of  the difference between natural frequencies: 
    $ \frac{\Delta \omega}{\omega} \equiv \frac{\omega_{1} - \omega_{2}}{\omega} $. }
    \label{fig:delta-omega}
  \end{center}
\end{figure}

\end{document}